\documentclass[a4,12pt,amsmath]{amsart}
\usepackage{amssymb,epic,eepic,epsfig,amsbsy,amsmath}
\usepackage{amscd}
\usepackage{times}
\usepackage[headings]{fullpage}
\numberwithin{equation}{section}

\newtheorem{thm}{Theorem}[section]
\newtheorem{defn}[thm]{Definition}
\newtheorem{cor}[thm]{Corollary}
\newtheorem{prop}[thm]{Proposition}
\newtheorem{lem}[thm]{Lemma}
\newtheorem{rem}[thm]{Remark}

\def\FFT{\mathop{\mathrm {FFT}}\nolimits}
\def\add{\mathop{\mathrm {add}}\nolimits}
\def\inverse{\mathop{\mathrm {inverse}}\nolimits}
\def\H{\mathop{\mathrm {H}}\nolimits}
\def\U{\mathop{\mathrm {U}}\nolimits}

\begin{document}

\title[Accounting Principles are Simulated on Quantum Computers]{Accounting Principles are Simulated on Quantum Computers}
\author[Do Ngoc Diep and Do Hoang Giang]{Do Ngoc Diep${}^1$ and Do Hoang Giang${}^2$}
\maketitle
%\begin{slide}
\begin{abstract} The paper is devoted to a new idea  of simulation
of accounting by quantum computing.
We expose the actual accounting principles in a  pure mathematics
language. After that we simulated the accounting principles on
quantum computers. We show that all arbitrary accounting actions
are exhausted by the described basic actions.  The main problem of
accounting are reduced to some system of linear equations in the
economic model of Leontief. In this simulation we use our
constructed quantum Gau\ss-Jordan Elimination to solve the problem
and the time of quantum computing  is some square root order
faster than the time  in classical computing.
\end{abstract}

\section{Introduction}
It is clear that accounting is the language of business, more
precisely {\it accounting is an information system that
identifies, records, and communicates the economic events of an
organization to interested users. The main economic events are
recorded and then processed by the own rules of accounting in
order to give adequate informations to managers and users.}

There is some basic accounting equation
\begin{center} \fbox{assets= Liabilities + Owner's Equity} \end{center}
or the extended basic accounting equation
\begin{center}assets(Cash + Account Receivables + Supplies + Equipments+ ...)\end{center}

\begin{center} = Liabilities (Account Payable + ...) + Owner's Equity(Capital + Revenue -
Expenses -Drawing + ...) \end{center}

We observe that {\it all recording accounts and business actions
are vectors and linear operators in linear algebra}. We construct
in this paper the linear algebra model for accounting, in
particular we define accounts as some 2-component vectors (one
component for debit and one for credit).

The problems of accounting are simulated on computer programs like
Excel, etc...  The next essential observation is that we {\it
could simulate accounting  by quantum computing} and therefore we
use the new technology of quantum computers.

The paper is organized as follows. In the next section we describe
accounting as some pure mathematical theory. In Section 4 we
recall some basic of quantum computers and quantum algorithms. In
Section 4 we prove how we can simulate accounting on quantum
computers and in Section 5 we so the solution to the economic
problem in leontief modelling. We refer the readers to
\cite{recipes} some program for Gau\ss-Jordan Elimination in C++
language.

\section{Mathematical Accounting}
It is well-known that {\it accounting} is the information system
that identifies, records and communicates the economic events of
an organization to interested users. In some ordinary language
people records these events by {\it transaction}, i.e. the
economic events of an enterprize that are recorded by {\it
accountants}. Accountants then {\it journalize} the transactions
into some account, normally in form of some T-accounts or
multiple-column accounts. A {\it T-account} is a quantity that has
some name for reference, some amount (number) that should be
putted in some {\it debit} or {\it credit} side. The amount is
normally some amount of money or equivalents, that we can add some
of them together or multiply with some scalar. This notion can be
expressed in a purely mathematical language as follows.
\begin{defn}
{\rm A {\it T-account} \fbox{$x\vert y$} is a real two dimensional
vector space with a fixed standard basis consisting of two
vectors: $|0\rangle:=\fbox{$1|0$}$ for debit side and $|1\rangle:=
\fbox{$0|1$}$ for credit side. Any vector of form $x |0\rangle +
y|1\rangle$ represents an {\it amount $x$ on debit side} and an
{\it amount $y$ on the credit side}. }\end{defn}

One transaction can be journalized as some subspace in a sum of
T-accounts,  corresponding to that action.
\begin{cor}
A transaction can be presented as a  subspace in the vector space,
generated by the corresponding T-accounts.
\end{cor}

\begin{cor}
A  journal can be presented as  the vector space, generated  by
the corresponding T-accounts, i.e. a vector space of form
\begin{equation}(\mathbf R^2)^k =\bigoplus_k(\mathbf R^2):=
\underbrace{\mathbf R^2 \oplus \dots \oplus \mathbf R^2}_{\mbox{k
times}}.\end{equation}
\end{cor}

\begin{cor}
A {\rm special journal} can be presented as   the vector space,
generated by the corresponding T-accounts, i. e. a vector space of
form \begin{equation}(\mathbf R^2)^k =\oplus_k(\mathbf R^2):=
\underbrace{\mathbf R^2 \oplus \dots \oplus \mathbf R^2}_{\mbox{k
times}}.\end{equation}
\end{cor}

After journalized transactions, accountants post the  {\it journal
entries}  to {\it special ledger} or {\it general ledger} or {\it
subsidiary ledger}. This process extract the actions in some
specific groups of action concerning some debit or credit amounts.

\begin{cor}
A {\rm subsidiary ledger} or {\rm general ledger} can be presented
as   a vector subspace of the vector space, generated by the
corresponding T-accounts, i. e. a subspace of $(\mathbf R^2)^k$.
\end{cor}

\subsection{Basic accounting equation}
In accounting the basic accounts like: Assets (A), Liabilities
(L), Owner's Equity (OE): Capital (C), Revenue (R), Expenses (E)
and Drawing (D) are related in the so called {\it the basic
accounting equation} \begin{equation} A = L + OE\end{equation} or
the {\it extended basic accounting equation}
\begin{equation} A = L + R - E + C - D.\end{equation}

\begin{prop}
The basic accounting equations and the extended accounting
equations are  systems of linear equations on the vector space of
all accounts.
\end{prop}
{\sc Proof.} Indeed each of the accounts is a 2-dimensional real
vector  space. And the equations are written for each debit or
credit side. \hfill$\Box$

\subsection{Worksheet}
In accounting a {\it worksheet} is a table of accounts with the
pair of  columns representing debit and credit sides of: {\it
Trial Balance, Adjustments, Adjusted Trial Balance, Income
Statements, and Balance Sheet}
\begin{prop}
Each of the quantities: Trial Balance, Adjustments, Adjusted Trial
Balance, Income Statements,  and Balance Sheet can be represented
by a vector subspace. A worksheet is a sum of such the vector
subspaces.
\end{prop}
{\sc Proof.} The table  in the worksheet is exactly the sum of
columns. Each of column represents a debit or credit of some
account. \hfill$\Box$

We fix a standard basis  for each of the listed subspaces and
denote the vector representing this worksheet as a column $X$.

\subsection{Business Action}

At the end of a {\it accounting cycle}  accountants {\it transfer
the net income} to the capital for the next accounting cycle. In
our model we have therefore the following statement.
\begin{prop}
The process of  transferring an amount from one place to another
place in the worksheet is some linear operator in the worksheet
vector space.
\end{prop}
Indeed we have often exactly the same quantities at the initial
place  and at the targeting places.

\begin{cor} For a business action, there is a linear map and its
corresponding  matrix $A$ such that $X$ is the input worksheet and
$AX$ is the output worksheet.
\end{cor}

\subsection{Leontief Models}
We consider in the economics the models of  Leontief: in one case
- the closed model when the output equal to the input, or in
another words, the consumption equals to the production and in
another case - the open model when the a part of production is
consumed by the producers and the another part of production is
consumed by external bodies.

\subsubsection{Closed Leontief Model AX=X}
In this model we have an worksheet $X$ as the input vector and a
business operation, represented by a matrix $A$, the output vector
is $AX$. The equation of the model is \begin{equation}AX =
X.\end{equation} The equation is written in equivalent form of a
homogeneous system
\begin{equation}(I-A)X=0.\end{equation} The following result is
well-known in economics.
\begin{thm}
If the entries of the output-input matrix $A$ are positive and if
the the  sum of each column of $A$ equal 1, then this system has
one-parameter solution.
\end{thm}
Remark that the solution can be obtained by the well-known
Gau\ss-Jordan  Elimination procedure.

\subsubsection{Open Leontief Model AX+D=X}
In the open model the vector $D$ is called the {\it demand vector}
We can rewrite the system in form \begin{equation}(I-A)X =
D.\end{equation} It is also well-known the following result.
\begin{thm}
If the entries of the output-input  matrix $A$ are positive and if
the the sum of each column of $A$ less than 1, then the matrix
$I-A$ is invertible and this system has a unique  solution
\begin{equation}X = (I-A)^{-1}D.\end{equation}
\end{thm}

Remark once again that the solution can be obtained by  the
well-known Gau\ss-Jordan Elimination procedure.

In actual accounting, there are some computer  programs for
solving equation in Excel, what we often use. If the size of
matrix $A$ is $n \times n$ then in classical computation estimate
of $2^n$ computations. In the rest of this paper we propose
another simulation of accounting by quantum computing, which let
to solve system faster, estimated of order $2^{n/2}$.

\subsection{Complexification}
In the previous model, debit and credit  are deterministic
quantities. In practice, there are some case, especially in
finance and stock-markets, we do use stochastic model. In that
case we extend the model of T-account from $\mathbf R^2$ to
complex plane $\mathbf C^2$.

\begin{defn}{\rm
A state of a T- account is a vector with  complex components
\begin{equation}\alpha|0\rangle + \beta|1\rangle\in \mathbf C^2,\end{equation} where
\begin{equation}|\alpha|^2 + |\beta|^2 =1.\end{equation} The number $|\alpha|^2$ is the
probability we have an amount in debit side. }\end{defn}

\begin{cor}
An amount \begin{equation}x\alpha|0\rangle +
y\beta|1\rangle\end{equation} in a $T$-account is a distribution
that we have amount $x$ on debit side with probability
$|\alpha|^2$ and have amount $y$ on credit side with probability
$|\beta|^2$.
\end{cor}

\section{Quantum Computers}
In this section we give a brief review of quantum computers, as
we need in the rest of this paper.

\subsection{Qubits}
A fundamental notion of quantum computers is qubit.
\begin{defn}{\rm
A {\it qubit} is a quantum system the Hilbert space of its quantum
states  is the 2-dimensional complex plane $\mathbf C^2$, and
therefore a quantum state of a qubit is a normalized complex
vector $\mathbf v\in \mathbf C^2$
\begin{equation}\mathbf v = \left[\begin{array}{c} v_0\\ v_1\end{array}\right],
||\mathbf v|| = |v_0|^2 + |v_1|^2 = 1.\end{equation}
}\end{defn}

\begin{lem}
Every $n$-digit integer  number can be written as a tensor product
of $n$ qubits \begin{equation}a = a_0 \otimes \dots \otimes
a_{n-1} \in (\mathbf C^2)^{\otimes n}, a_i \in \mathbf F_2=
\mathbf Z/2\mathbf Z.\end{equation}
\end{lem}
{\sc Proof.} Every integer can be written in binary form
\begin{equation}a= a_0 + a_12 + \dots + a_n2^n, a_i \in \mathbf F_2
=\{ 0, 1 \}.\end{equation}
Therefore we have $|a_0\rangle \otimes \dots \otimes |a_n\rangle
\in (\mathbf C^2)^n.$ \hfill$\Box$

Let us consider the additive groups $G= (\mathbf F_2)^n$.  A
unitary character is a continuous homomorphism $\chi: G \to
\mathbf S^1 = \mathbf U(\mathbf C)$ from the group $G$ into the
group of unitary automorphism of $\mathbf C$.
\begin{lem}
Every (multiplicative) unitary character of the additive groups
$G=(\mathbf F_2)^n$ if of the form
\begin{equation}\chi_a(x) = \exp\left({\frac{2\pi i}{2^n}\sum_{i=1}^n a_ix_i}\right).\end{equation}
\end{lem}
{\sc Proof.} The lemma is an easy exercise from the group
representation  theory.\hfill$\Box$

It is convenient to write the number in fractional form, i.e.
write
\begin{equation}0.a = \frac{a_0}{q^n} + \frac{a_1}{q^{n-1}} + \dots .\end{equation}
The integers modulo $2^n$ can be written as
\begin{equation}\bar{k} \mapsto e^{2\pi i 0.k}, \mbox{ where } 0.k = k/q^n,\end{equation}
for $k=0,1,\dots, q^n-1$. Let us consider the Haar measure $\mu(k)
= \frac{k}{q^n}$. Then we have the natural {\it Fast Fourier
Transform} (FFT).
\begin{defn}{\rm
The transformation \begin{equation}\FFT : |y\rangle \mapsto
\frac{1}{2^n}\sum_{k=0}^{q^n-1} e^{2\pi i
y_k\frac{k}{q^n}}|k\rangle\end{equation} is called the {\it Fast
Fourier Transform} of $|y\rangle$ }\end{defn} After that we can
formally write a number $|y\rangle$ in the form of $\exp(2\pi i
0.y_{n}y_{n-1}\dots y_0)$, i.e. we have
\begin{equation}\FFT : |y\rangle \mapsto
\exp(2\pi i 0.y_{n}y_{n-1}\dots y_0) \label{ffff} \end{equation}
\subsection{Quantum addition}
By using FFT we can write integers in the form \ref{ffff} i.e. we
have a 1-1 correspondence
\begin{equation}\CD a @>\FFT>> e^{2\pi i 0.a_{n}a_{n-1}\dots a_0}\endCD\end{equation}
\begin{equation}\CD b @>\FFT>> e^{2\pi i 0.b_{n}b_{n-1}\dots b_0}\endCD\end{equation}

\begin{defn}
The sum of two numbers $a$ and $b$ is defined as follows.
\begin{equation}\CD a @>\FFT>> e^{2\pi i0.a_n\dots a_0} @> \add\;  0.0\dots 0b_0 >>
e^{2\pi i0.a_na_{n-a}\dots a_1(a_0+b_0)} @>\add\; 0.0\dots 0b_1
>> \dots\endCD\end{equation}
\begin{equation}\CD \dots @> \add \; 0.b_n >> e^{2\pi i 0.(a_n+b_n)\dots
(a_0+b_0)} @>\FFT \inverse>> a+b\endCD \end{equation}
\end{defn}

\begin{rem} The reduction (memorize $[(a_i+b_i)/q]$ to
$a_{i+1} + b_{i+1}$ is the phase transition with the matrix
\begin{equation}\left[\begin{array}{cc}
1 & 0\\
0 & e^{2\pi i\frac{[(a_i+b_i)/q]}{q^{i-1}}}
\end{array}\right]\end{equation}
\end{rem}

\subsection{Quantum gates}
Let us recall some fundamental gates in quantum computing.

\subsubsection{Hadamard gate}
\begin{defn} The 1-qubit Hadamard gate $-\fbox{$\H$}-$ {\rm is defined by the
matrix}
\begin{equation}\H = \frac{1}{\sqrt{2}}\left[\begin{array}{cc}
1 & 1\\
1 & -1 \end{array}\right].\end{equation}
\end{defn}
{\rm The Hadamard gates acts on 1-qubits as follows}.
\begin{equation}\CD|x\rangle @>-\fbox{$\H$}->> \frac{1}{\sqrt{2}}
\left((-1)^x|x\rangle + |1-x\rangle\right).\endCD\end{equation}

\subsubsection{Phase gate}
\begin{defn} The phase 2-qubit gate $-\fbox{$\Phi$}-$ is defined by the matrix
\begin{equation}\Phi = \frac{1}{\sqrt{2}}\left[\begin{array}{cc}
1 & 0\\
0 & e^{i\Phi}\end{array}\right].\end{equation} it produces an
action
\begin{equation}\CD |x\rangle @>-\fbox{$\Phi$}->> \frac{1}
{\sqrt{2}}e^{i\Phi}|x\rangle.\endCD\end{equation}
\end{defn}

\subsubsection{CNOT(XOR) gate}
\begin{defn}{\rm
The 2-qubit CNOT (XOR)} gate {\rm is defined by the matrix}
\begin{equation}C = \left[\begin{array}{cccc}
1 & 0 & 0 & 0\\
0 & 1 & 0 & 0\\
0 & 0 & 0 & 1\\
0 & 0 & 1 & 0\end{array}\right].\end{equation} {\rm The action of
this gate is given by}
\begin{equation}\left\{\begin{array}{rcl}
|x\rangle & \to & |x\rangle\\
|y\rangle & \to & |x \oplus
y\rangle,\end{array}\right.\end{equation} {\rm where} $x\oplus y
:= x+ y \mod(2).$ \end{defn}

\subsubsection{Unitary gates}
\begin{defn}
The unitary gate -\fbox{$\U$}- {\rm is defined by a unitary matrix
U},
\begin{equation}\mathbf \U = \left[\begin{array}{cc}
1 & 0\\
0 & \U\end{array}\right].\end{equation}
\end{defn}

\subsubsection{Problem of quantum computing}
One of the major result concerning this problem is the following
classification result.
\begin{thm}[Universal gates]
One-qubit gates, CNOT gate, phase gate and one of unitary gate
generate all the other gates.
\end{thm}

\section{Quantum Grover's Search Algorithm}
\subsubsection{Black-box (Query) problem}
The problem is to compute some Boolean function of type
\begin{equation}f: \{0,1\}^n \to \{0,1\}.\end{equation}

\subsection{Deutsch's original problem}
The problem is to check whether the above defined Boolean function
is {\it balance}: the number of values 0's equals to the numbers
of values 1's, or {\it constant}.

\subsection{Search problem}
Given a Boolean function $f: \{0,1\}^n \to \{0,1\},$ defined by
$f_k(x) = \delta_{xk}.$ Find $k$.

\subsection{Deutsch's scheme $-\fbox{H}-\fbox{f}-\fbox{H}-$}
The Deutsch's scheme produces the effects (modulo a constant, we
often omit it in order to not the formula longer, but it is easily
recovered in the case of necessary):
\begin{equation}\CD |x\rangle(|0\rangle -|1\rangle) @>-\fbox{H}-
\fbox{f}-\fbox{H}->> \underbrace{(-1)^{f(x)}|x\rangle}_{\mbox{measurement}}
(|0\rangle-|1\rangle).\endCD\end{equation}
In this scheme, the second qubit is of no interest, while the
first qubit has state \begin{equation}(-1)^{f(0)}|0\rangle +
(-1)^{f(1)}|1\rangle = \left\{\begin{array}{ll}
\pm(|0\rangle + |1\rangle), & \mbox{if } f=const\\
\pm(|0\rangle - |1\rangle), & \mbox{if } f=\mbox{\it
balanced}\end{array}\right.\end{equation}

\subsection{Quantum Grover's Search Algorithm}
Scheme: $-\fbox{$f_k$}-\fbox{H}-\fbox{$f_0$}-\fbox{H}-$.

This algorithm has special effect: Choose the state
\begin{equation}|S\rangle = \frac{1}{2^{N/2}}\sum_{i=1}^{N-1}
|i\rangle,\end{equation}
and produce the Grover's iterate
\begin{equation}\CD |\psi\rangle(|0\rangle-|1\rangle)
@>-\fbox{$f_k$}-\fbox{H}-\fbox{$f_0$}-\fbox{H}->>
|\mbox{measurement}\rangle (|0\rangle -|1\rangle).\endCD\end{equation}

This iterate does nothing to any basis element $|i\rangle$ except
for $|k\rangle$ is changing into $-|k\rangle$, i.e. {\it the
reflection in the hyperplane perpendicular to the plane}
\begin{equation}\H|0\rangle =
|S\rangle =
\frac{1}{2^{N/2}}\sum_{i=1}^{N-1}|i\rangle,\end{equation} i.e.
rotation in the plane generate by $|k\rangle$ and $|S\rangle$ an
the Grover's iterate is a rotation of twice the angle from
$|k^\perp\rangle$ to $S^\perp\rangle$.

\section{Simulation}

In accounting, see \cite{wkk} the normal balance of an account  is
on the side where an increase in account is recorded. {\it Trial
balance} of an account is the maximum part of amounts that balance
debit and credit sides of that account. In general {\it trial
balance} is a list of accounts and their balances at a given time.
\begin{thm} The decomposition of an T-account into trial balance
and normal balance can be obtained by applying the 2-qubit CNOT
gate to that decomposed account into the sum of two accounts.
\end{thm}
{\sc Proof.} Suppose that the $T$-account is of the form
\fbox{$a|b$},  i. e. has amount $a$ on the debit side and amount
$b$ on the credit side. If $a>b$ we have a debit normal balance
$a-b$. This means that \begin{equation}\mbox{\fbox{$a\vert b$} =
(a-b)\fbox{$1\vert 0$} + b \fbox{$1\vert 1$}, if  $a\geq
b$}\end{equation} and if $a \leq b$, we have a credit normal
balance $b-a$ and \begin{equation}\mbox{\fbox{$a\vert b$} =
(b-a)\fbox{$0\vert 1$} + a \fbox{$1\vert 1$}, if $a\leq
b$.}\end{equation} These operations are well provided on quantum
computers. \hfill$\Box$

In business and accounting, we often need to move one amount from
one account to another, for examples at the end of a accounting
cycle, all temporary accounts should be closed and move net income
to the capital account of the next accounting cycle.
\begin{thm}
Transferring an amount  from one account to another account can be
simulated by the rule of tensor product of accounts.
\end{thm}
{\sc Proof.} In quantum computers,  the 2-dimensional vector
spaces, representing accounts, are  multiplied as tensor products.
The coefficient of one factor is transferred to the factor of
another account vector spaces.
\begin{equation}\dots\otimes|kv_i\rangle\otimes \dots\otimes
|v_j\rangle\otimes\dots =
  \dots\otimes|v_i\rangle\otimes \dots\otimes |kv_j\rangle\otimes\dots
  .\end{equation}
This simulates the process of transferring an amount from one account to another.
\hfill$\Box$

In {\it accrual-basis accounting } transactions that change a
company's  financial statements are recorded in the period in
which the events occurred, see \cite{wkk}. There are four types of
accrued accounts: prepaid expenses, unearned revenues, accrued
revenues,   accrued expenses. {\it Prepaid expenses} are the
expenses paid in cash and recorded as assets before they are used
for consumed. {\it Unearned revenues} are revenues received in
cash and recorded as liabilities and recorded before revenues are
earned. {\it Accrued expenses} are the expenses incurred but not
yet paid in cash or recorded. {\it Accrued revenues} are the
revenues earned but not yet received in cash or recorded. For
these accounts we need some adjustments by adding some adjustment
accounts. {\it Adjustment entries} are made at the end of an
accounting period to ensure that the accrued accounts are
recognized and matching the accounting principles.

\begin{thm}
The adjustment process in accounting can be obtained by applying
a 2-qubit C-NOT gate.
\end{thm}
{\sc Proof.} An adjustment consists of decomposing a normal
balance into a sum of two sub-amounts.  Then keep one components
(for incurred expenses, ....) and rotate the second component into
opposite side (from debit to credit side). In mathematical
language, we have \begin{equation}\mbox{\fbox{$a| b$} =
\fbox{$a_1|b$} + \fbox{$(a-a_1)|b$}}\end{equation} or
\begin{equation}\mbox{\fbox{$a| b$} = \fbox{$a|b_1$} +
\fbox{$a|(b-b_1)$}.}\end{equation} Twist one account and keep
another account is simulated be a 2-qubit C-NOT gate.
\hfill$\square$

\begin{thm}
In complexified accounting (i.e. stochastic accounting) the
rotation  of an account into the angle of $\frac{\pi}{4}$ is
simulated by Hadamard gate and flip between amounts in the credit
and debit sides.
\end{thm}
{\sc Proof.} The matrix of the Hadamard gate is decomposed into
the product of the matrix of flip the basis vectors and the
rotation matrix: \begin{equation}H=
\frac{1}{\sqrt{2}}\left[\begin{array}{cc} 1 & 1\\ 1 &
-1\end{array}\right] = \frac{1}{\sqrt{2}}\ \left[\begin{array}{cc}
1 & 1\\ -1 & 1
\end{array}\right].\left[\begin{array}{cc} 0 & 1\\ 1 & 0
\end{array}\right] .\end{equation} \hfill$\Box$
\begin{cor}
An arbitrary business action can be reduced to repeated actions of
the  previous kind: trial balance, transfer, closing, etc...
\end{cor}
{\sc Proof.} It is a corollary from the theorem of classification of gates.
\hfill$\Box$

\section{Quantum Gau\ss-Jordan Elimination Procedure}
We reduced solution of economic problem to solution of a linear
system of  equations. In this section we use quantum computing
technique to find that solution. The method is much more faster
than using classical computer technique.

Let is recall the main result of \cite{diepgiang}. We consider
the linear system of $N$ equations on $N$ variables
$$AX=\mathbf b.$$

{\bf Quantum Gau\ss-Jordan Elimination Algorithm:}
\begin{enumerate}
\item[Step 1] Use the Grover's Search algorithm to find out  the
first  non-zero $a_{i1}\ne 0$. \item[Step 2] If the search is
successful, produce the first leading $\mathbf 1$ in the first
place as $a_{11}$, else change to the next column and repeat step
1. \item[Step 3] Eliminate all other entries $a_{1,1},\dots,
a_{N,1}$ in the column. \item[Step 4] Change $N$ to $N-1$, control
if still $N>0$, repeat the procedure from the step 1. \item[Step
5] In backward eliminate all $a_{N-1,N}, \dots, a_{1,N}$.
\item[Step 6] Check if $N>0$, change $N$ to $N-1$ and repeat the
step 5.
\end{enumerate}
\begin{thm}
In the Quantum Gau\ss-Jordan Elimination Algorithm one needs at
most
\begin{equation}\frac{N(N-1)(2N+1)}{3} + \left[\sqrt{2}\frac{(\sqrt{2})^N -1}{\sqrt{2}-1}\right] \sim O(2^{N/2}) \end{equation}
operations.
\end{thm}
{\sc Proof.} See \cite{diepgiang} \hfill$\Box$

{\parindent=0pt \sc ${}^1$ Institute of Mathematics,
Vietnam Academy of Sciences and Technology, 18 Hoang Quoc Viet Road,
Cau Giay District, 10307 Hanoi, Vietnam\\
{\tt Email: dndiep@math.ac.vn}\\
{\rm and}\\
${}^2$ K47A1T, Department of Mathematics, Mechanics and Informatics,
College of Natural Sciences, Vietnam National University,
40 Nguyen Trai, Thanh Xuan District, Hanoi Vietnam\\
{\tt Email: dhgiang84@gmail.com}}

\begin{thebibliography}{llllllllll}
\bibitem[DG]{diepgiang} {\sc  D.N.Diep and D. H. Giang},  {\it
Quantum Gau\ss\ Jordan Elimination}, arXiv.org: math.QA/0511145

\bibitem[D1]{diep1}{\sc D. N. Diep}, {\it Quantum computers
and related mathematical structures}, J. of Mathematics Applications,
in Vietnamese, {\bf 2}(2004), No 1, 79-94.

\bibitem[PTVF]{recipes} {\sc W. H. Press, S. A. Teukolsky W. T. Velterling
and B. P. Flannery}, {\it Numerical Recipes in C}, Cambridge University Press,

\bibitem[WKK]{wkk}{\sc J. Weygandt, D. kieso and P. Kimmel},
{\it Accounting Principles}, 6th edition, J. Wiley \& sons, Inc.,
pp. 1138, U.S.A.,2002.
\end{thebibliography}
\end{document}